# Dropwise Condensation on Hydrophobic Cylinders

Kyoo-Chul Park[1,2], David Fox[1], Michelle Hoang[1], Brendan McManus[1], Joanna Aizenberg[1,2,3]

[1] John A. Paulson School of Engineering and Applied Sciences, Harvard University, Cambridge, MA, USA; [2] Wyss Institute for Biologically Inspired Engineering, Harvard University, Cambridge, MA, USA; [3] Department of Chemistry and Chemical Biology, Harvard University, Cambridge, MA, USA

**Abstract**

In this work, we studied the effect of the diameter of horizontal hydrophobic cylinders on droplet growth. We postulate that the concentration gradient created by natural convection around a horizontal circular cylinder is related to the droplet growth on the cylinder by condensation. We derive a simple scaling law of droplet growth and compare it with experimental results. The predicted negative exponent of drop diameter ($d$) as a function of cylinder diameter ($D$) at different time points is similar to the general trend of experimental data. Further, this effect of cylinder diameter on droplet growth is observed to be stronger than the supersaturation conditions created by different surface temperatures.

**Main Text**

Controlling droplet growth is fundamentally important for a variety of applications ranging from various heat exchangers to water harvesting systems[1-5]. Facilitating droplet growth to faster reach critical droplet diameters ($d$) for shedding or jumping in the same condensation environment (*e.g.*, surface temperature, relative humidity, humid air temperature) can lead to saving energy required to achieve the designed performance of such phase change heat transfer systems[3-11] or to maximize the phase change heat transfer efficiency. To achieve faster drop growth, previous studies have mostly used hybrid surface wettability patterns[12,13]. Recently, the effect of convex surface topography on local preferential condensation by focused diffusion flux have been reported[11,14-16]. However, while biological species such as cacti and spider webs have employed a variety of geometries including cylindrical structures for condensation[17-19], studies on the effect of cylindrical topography itself on drop growth are rare[8,20].

In this work, we studied droplet growth on horizontally positioned cylinders because it is frequently found in natural water harvesting examples and widely used in phase change heat transfer systems. The surface temperature in the middle of the cylinders was kept at 6.5 ± 0.5°C (for Figs. 1-3) and monitored during the condensation, which is below the saturation point kept by relative humidity of 60 ± 5 % and atmospheric temperature of 23 ± 2°C. We used aluminum wires for two smaller diameter ($D$ = 1.02, 3.20 mm) cylinders. For the largest diameter ($D$ = 9.53 mm) cylindrical aluminum pipe, both ends were closed to minimize condensation inside the pipe and the surface temperatures at different polar angle positions were measured to be identical. The surface of each sample was first roughened using the same sandpaper (Sanding Sheet for Aluminum, 240 Grit, McMaster-Carr) to produce similar level of microscale surface roughness. The cylinders were then immersed in a 1 wt% solution of fluoroaliphatic phosphate ester fluorosurfactant (FS100, Mason Chemical Company) in 95:5 ethanol:water. The resultant hydrophobic microtextured surfaces did not produce jumping droplets during condensation experiments. Experiments (for Figs. 2, 3) were repeated more than 4 times and similar droplet growth on the same sample was observed each time, which confirmed that the chemical coating was not noticeably degraded during the experiments.

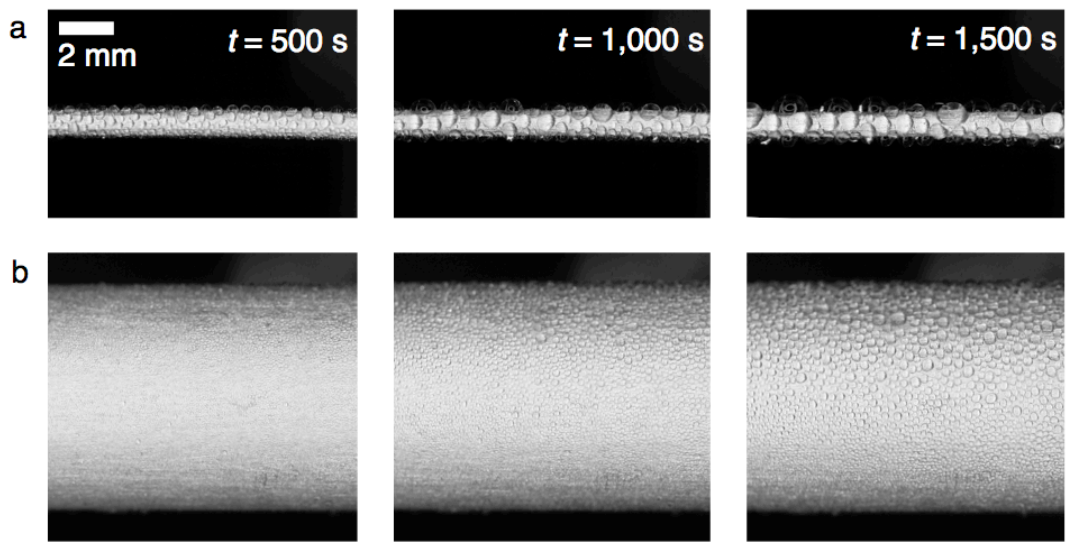

**Figure 1**. Dropwise condensation on rough hydrophobic cylinders of different diameter $D$. (a) $D$ = 1.02 mm and (b) $D$ = 9.53 mm.

Fig. 1 shows the time-lapsed images of droplets condensed on two cylinders of different diameters. The smaller diameter cylinder exhibits larger droplets across all the time points, indicating faster droplet growth, even though the roughness, temperature, relative humidity, and surface chemistry are carefully maintained to be identical. Although previous studies investigated various droplet-cylindrical structure interactions, this faster droplet growth on a smaller diameter cylinder has not been analyzed systematically[8,17-20].

To quantitatively analyze the effect of cylinder diameter on the droplet growth, we plotted the average diameter of droplets ($N = 5$) at the polar angle $\psi = 90°$ as a function of time (Fig. 2). We counted only the first generation droplets[15,16]. The log-log plot shows that all the three sets of data have similar exponents close to unity, which is generally observed in droplet growth plots[7,8,14-16]. However, the y-axis intercept at $t = 250$ s is different, similar to previous observations on different convex topographies[11].

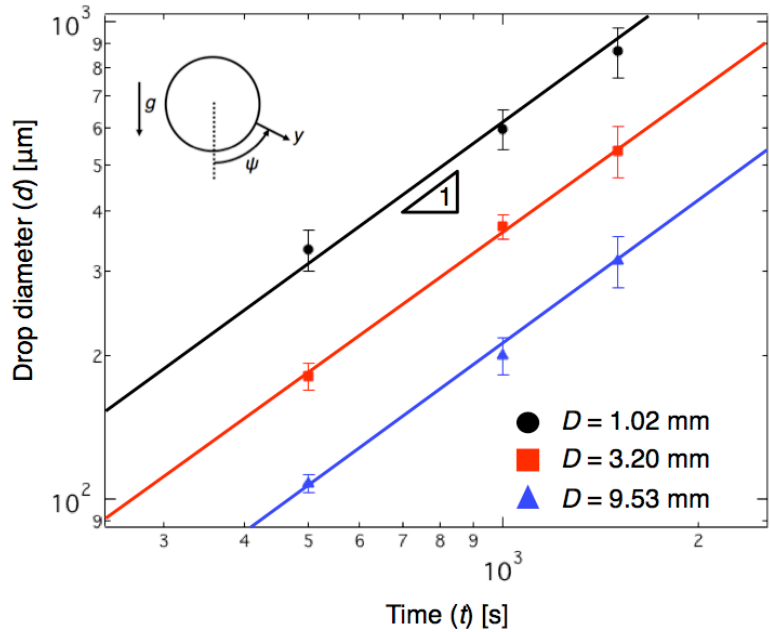

**Figure 2**. Quantitative analysis of the droplet growth on cylinders of different diameter as a function of time. All error bars are 1 s.d.

We postulate that this interesting phenomenon is related to diffusion flux[15] because all other components of the following governing equation[15] have been kept the same in all the experiments, similar to previous studies on different convex topographies[11,14],

$$\langle d \rangle = \phi_v \, t$$

where $\langle d \rangle$ is the average droplet diameter, $\phi_v$ is the diffusion flux, and $t$ is the time elapsed from the beginning of each condensation experiment.

To derive the relation between the $\phi_v$ and cylinder diameter ($D$), we checked previous numerical calculation study on the concentration gradient ($dC/dy$) created by natural convection generated by complex heat and mass transfer[21]. Assuming the same buoyancy effect generated by difference of humid air density around different diameter cylinders, as well as constant values of Prandtl number and Schmidt number ($Pr \sim 0.7$ and $Sc \sim 0.7$), the dependence of concentration gradient on the diameter of horizontal cylinders is derived as follows[21],

$$dC/dy \,|_{\psi=90°} \sim (D^{-0.25}) \, dC/d\eta$$

where $\eta$ is dimensionless coordinate normal to the cylindrical surface[21] and $dC/d\eta$ is the same regardless of $D$ and approximately linear when $\eta \leq 2$.

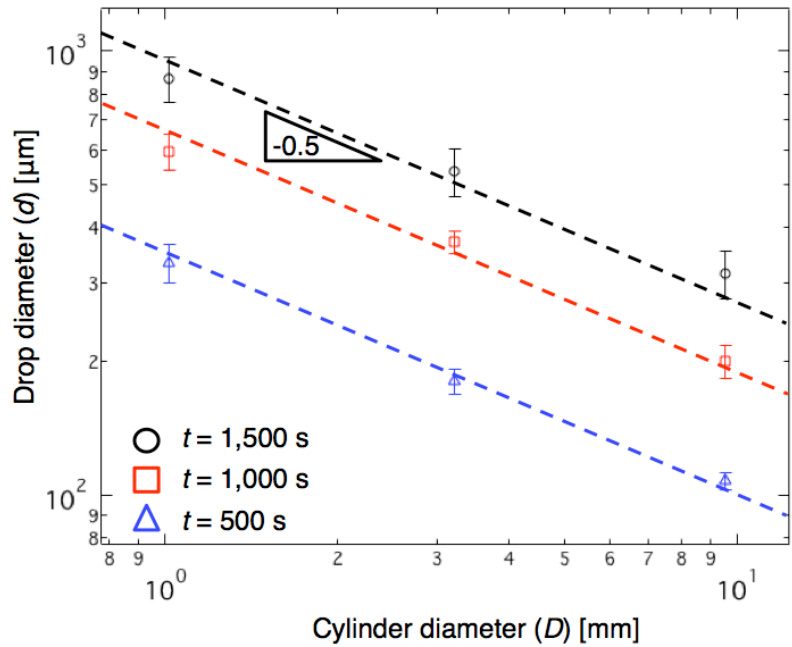

**Figure 3**. Dependence of droplet size on cylinder diameter at different time points. All error bars are 1 s.d.

By substituting this relation into the governing equation of diffusion (*i.e.,* Fick's Law),

$$\phi_v = D_{12} dC/dy \sim D^{-0.25}$$

where $D_{12}$ is the diffusion coefficient of water vapor in air at atmospheric pressure. This derived relation based on our hypothesis on the diffusion flux of water vapor explains the general trend of faster droplet growth on a smaller diameter cylinder - a negative exponent of the three droplet diameter - cylinder diameter curves observed in Fig. 3. However, the absolute value of the exponent in Fig. 3 is greater than -0.25 that is derived from Fick's Law. One of the reasons would be the stronger effect of coalescence of droplets between different polar angle positions when the cylinder diameter is smaller. For example, for $D = 1.02$ mm wire, drops are more likely to grow faster by the coalescence with the even larger droplets growing on the apex of the wire ($\psi = 180°$), compared to $D = 9.53$ mm cylinder. To understand this discrepancy between the exponent values from the derived relation and experiments, further experiments ruling out the effect of this coalescence on droplet growth should also be performed.

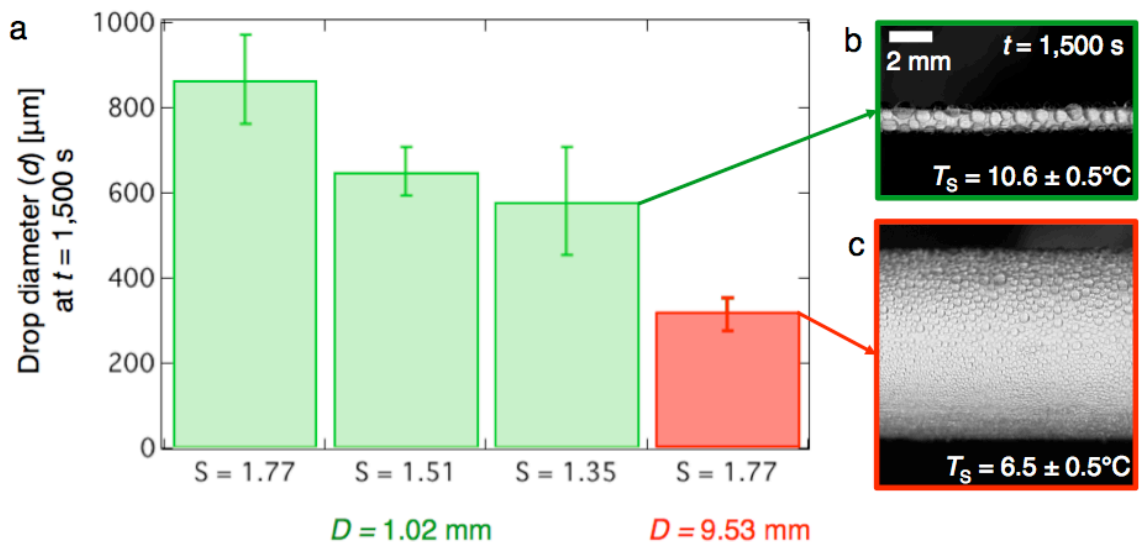

**Figure 4**. Fast droplet growth on a smaller diameter cylinder in less favorable surface temperature conditions. The last green bar ($S = 1.35$) and red bar ($S = 1.77$) in (a) correspond to images in (b) and (c), respectively. All error bars are 1 s.d.

The effect of cylinder diameter is further compared with the effect of supersaturation (*i.e.*, the ratio of the vapor pressure to the saturation pressure corresponding to the cylinder surface temperature ($S = P_V/P_S$)) created by different surface temperatures in the same humid air temperature and relative humidity environment in Fig. 4. We analyzed droplet growth on the smallest diameter wire ($D$ = 1.02 mm, Fig. 4a) of different surface temperatures compared to the largest diameter cylinder ($D$ = 9.53 mm, Fig. 4b) of the surface temperature $T_S$ = 6.5 ± 0.5°C. Fig. 4 shows that even in the higher, less favorable surface temperature conditions ($T_S$ = 10.6 ± 0.5°C), the droplet size on the smaller diameter wire ($S$ = 1.35) is greater than that on the largest diameter cylinder ($S$ = 1.77).

In conclusion, we studied the effect of cylinder diameter on droplet growth in condensation environments. To explain the faster droplet growth on a smaller cylinder diameter cylinder, we hypothesized that the concentration gradient generated by the natural convection around the horizontal circular cylinders is related to the droplet growth. We derived a simple scaling law including the exponent of -0.25 and compared this droplet diameter vs. cylinder diameter relation to experimental results. Further, this strong effect of cylinder diameter on droplet growth overcomes the unfavorable supersaturation condition created by higher surface temperatures. We envision that the effect of cylinder diameter on condensation could be applied to understand many different phase change phenomena including deposition and evaporation of various vapor molecules on cylindrical surfaces.

**References**

1. Clus, O., Ortega, P., Muselli, M., Milimouk, I. & Beysens, D. Study of dew water collection in humid tropical islands. *J. Hydrol.* **361**, 159–171 (2008).
2. Malik, F. T., Clement, R. M., Gethin, D. T., Krawszik, W. & Parker, A. R. Nature's moisture harvesters: a comparative review. *Bioinspir. Biomim.* **9**, 031002 (2014).
3. Miljkovic, N. et al. Jumping-droplet-enhanced condensation on scalable superhydrophobic nanostructured surfaces. *Nano Lett.* **13**, 179–187 (2013).
4. Rose, J. W. Dropwise condensation theory and experiment: a review. *Proc. Inst. Mech. Eng. A* **216**, 115–128 (2002).
5. Xiao, R., Miljkovic, N., Enright, R. & Wang, E. N. Immersion condensation on oil-infused heterogeneous surfaces for enhanced heat transfer. *Sci. Rep.* **3**, 1988 (2013).
6. Chen, C.-H. et. al. Dropwise condensation on superhydrophobic surfaces with two-tier roughness. *App. Phys. Lett.* **90**, 173108 (2007).
7. Boreyko, J. B. & Chen, C.-H. Self-propelled dropwise condensate on superhydrophobic surfaces. *Phy. Rev. Lett.* **103**, 184501 (2009).
8. Zhang, K. et. al. Self-propelled droplet removal from hydrophobic fiber-based coalescers. *Phy. Rev. Lett.* **115**, 074502 (2015).
9. Kim, P. et al. Liquid-infused nanostructured surfaces with extreme anti-ice and anti-frost performance. *ACS Nano* **6**, 6569–6577 (2012).
10. Anand, S., Paxson, A. T., Dhiman, R., Smith, J. D. & Varanasi, K. K. Enhanced condensation on lubricant-impregnated nanotextured surfaces. *ACS Nano* **6**, 10122–10129 (2012).
11. Park, K.-C. et. al. Condensation on slippery asymmetric bumps. *Nature* advance online publication, 24 February 2016 (DOI 10.1038/nature16956).
12. Varanasi, K. K., Hsu, M., Bhate, N., Yang, W. & Deng, T. Spatial control in the heterogeneous nucleation of water. *Appl. Phys. Lett.* **95**, 094101 (2009).

13. Mishchenko, L., Khan, M., Aizenberg, J. & Hatton, B. D. Spatial control of condensation and freezing on superhydrophobic surfaces with hydrophilic patches. *Adv. Funct. Mater.* **23**, 4577–4584 (2013).

14. Medici, M.-G., Mongruel, A., Royon, L. & Beysens, D. Edge effects on water droplet condensation. *Phys. Rev. E* **90**, 062403 (2014).

15. Viovy, J. L., Beysens, D. & Knobler, C. M. Scaling description for the growth of condensation patterns on surfaces. *Phys. Rev. A* **37**, 4965–4970 (1988).

16. Beysens, D. Dew nucleation and growth. *C. R. Phys.* **7**, 1082–1100 (2006).

17. Zheng, Y. et al. Directional water collection on wetted spider silk. *Nature* **463**, 640–643 (2010).

18. Malik, F. T. et al. Dew harvesting efficiency of four species of cacti. *Bioinspir. Biomim.* **10**, 036005 (2015).

19. Wang, Q., Yao, X., Liu, H., Quéré, D. & Jiang, L. Self-removal of condensed water on the legs of water striders. *Proc. Natl Acad. Sci. USA* **112**, 9247–9252 (2015).

20. Stricker, L. & Vollmer, J. Impact of microphysics on the growth of one-dimensional breath figures. *Phy. Rev. E* **92**, 042406 (2015).

21. Hasan, M. & Mujumdar, A. S. Laminar boundary-layer analysis of simultaneous mass and heat transfer in natural convection around a horizontal cylinder. *Energy Research* **11**, 359-371 (1987).